\documentclass[a4paper,superscriptaddress,twocolumn,prl]{revtex4}

\usepackage{amsmath}
\usepackage{amsfonts}
\usepackage{amssymb}
\usepackage{color}
\usepackage{units}
\usepackage{verbatim}
\usepackage{graphicx}

\newcommand{\beq}{\begin{equation}}
\newcommand{\eeq}{\end{equation}}
\newcommand{\bqa}{\begin{eqnarray}}
\newcommand{\eqa}{\end{eqnarray}}

\newcommand{\dg}{^\dagger}

\newcommand{\erf}[1]{Eq.~(\ref{#1})}

\newcommand{\bra}[1]{\left\langle{#1}\right|}
\newcommand{\ket}[1]{\left|{#1}\right\rangle}

\newcommand{\sch}{Schr\"odinger}

\newcommand{\cu}[1]{\left\{ {#1} \right\}}
\newcommand{\ro}[1]{\left( {#1} \right)}

\newcommand{\s}[1]{\hat\sigma_{#1}}

\newcommand{\Str}{\mathcal{S}}
\newcommand{\red}{\color{red}}

\newcommand{\rfrac}[2]{\left({#1}/{#2}\right)}

\begin{document}

\title{Experimental EPR-Steering of Bell-local States}

\author{D.\ J.\ Saunders}
 \affiliation{Centre for Quantum Dynamics, Griffith University, Brisbane 4111 Australia,}
  \affiliation{Centre for Quantum Computer Technology}
\author{S.\ J.\ Jones}%
 \affiliation{Centre for Quantum Dynamics, Griffith University, Brisbane 4111 Australia,}
  \affiliation{Centre for Quantum Computer Technology}
  \author{H.\ M.\ Wiseman\footnote{email: H.Wiseman@griffith.edu.au}}
 \affiliation{Centre for Quantum Dynamics, Griffith University, Brisbane 4111 Australia,}
  \affiliation{Centre for Quantum Computer Technology}
  \author{G.\ J.\ Pryde\footnote{email: G.Pryde@griffith.edu.au}}
 \affiliation{Centre for Quantum Dynamics, Griffith University, Brisbane 4111 Australia,}
  \affiliation{Centre for Quantum Computer Technology}

\begin{abstract}
Entanglement is the defining feature of quantum mechanics, and understanding the phenomenon is essential at the foundational level and for future progress in quantum technology. The concept of \textit{steering} was introduced in 1935 by \sch~\cite{SchPCP35} as a generalization of the Einstein-Podolsky-Rosen (EPR) paradox \cite{EinEtalPR35}. Surprisingly, it has only recently been formalized as a quantum information task with arbitrary bipartite states and measurements \cite{Wiseman2007,JonWisDoh07,CavJonWisRei09}, for which the existence of entanglement is necessary but not sufficient. Previous experiments in this area \cite{OuPereira1992,BowenSchnabel2003,HaldSorensen1999,HowellBennink2004} have been restricted to the approach of Reid \cite{ReiPRA89}, which followed the original EPR argument in considering only two different measurement settings per side.  Here we implement more than two settings so as to be able to demonstrate experimentally, for the first time, that EPR-steering occurs for mixed entangled states that are Bell-local (that is, which cannot possibly demonstrate Bell-nonlocality). Unlike the case of Bell inequalities \cite{Altepeter:2005,AciEtalPRA06,BrunGis08}, increasing the number of measurement settings beyond two---we use up to six---dramatically increases the robustness of the EPR-steering phenomenon to noise. 
\end{abstract}

\maketitle

EPR formulated their famous``paradox'' to highlight the (to them unacceptable) spooky-action-at-a-distance in the Copenhagen interpretation of quantum mechanics, where ``as a consequence of two different measurements performed upon the first [Alice's] system, the second [Bob's] system may be left in states with two different [kinds of] wave functions'' \cite{EinEtalPR35}. In EPR's specific example, the two different kinds of wave-functions were position and momentum eigenstates, which are clearly incompatible because ``precise knowledge of [$Q$] precludes such a knowledge of [$P$]'' \cite{EinEtalPR35}. In this paradigm, Reid \cite{ReiPRA89} first developed quantitative criteria for the experimental demonstration of the EPR paradox based on Heisenberg's uncertainty relation $(\Delta P)(\Delta Q) \geq \hbar /2$. However, in the same year as the EPR paper, \sch\ introduced the term {\em steering} \cite{SchPCP35} to describe the EPR paradox, and generalized it to more than two measurements, saying: ``Since I can predict either [$Q$] or [$P$] without interfering with [Bob's] system, \ldots [Bob's system] must know both answers; which is an amazing knowledge. [Bob's system] does not only know these two answers but a vast number of others.'' It is only very recently that general EPR-steering inequalities, allowing for measurements of an arbitrary number of different observables by Alice and Bob, have been developed \cite{CavJonWisRei09}, following the formal definition of steering \cite{Wiseman2007,JonWisDoh07}.  In this paper we exploit the more general formalism for the first time,  demonstrating EPR-steering with discrete binary-outcome measurements on Werner states \cite{WerPRA89} of a pair of photon-polarization qubits. This family of states is well studied, and it is proven that some of the states we use to demonstrate EPR-steering violate no Bell inequality. This is not the case for any of the states used in previous demonstrations \cite{OuPereira1992,BowenSchnabel2003,HaldSorensen1999,HowellBennink2004} of the EPR paradox, which relied upon the EPR-Reid inequalities, and used measurements with continuous outcomes.

\begin{figure}\begin{center}
\includegraphics[width=.94\linewidth]{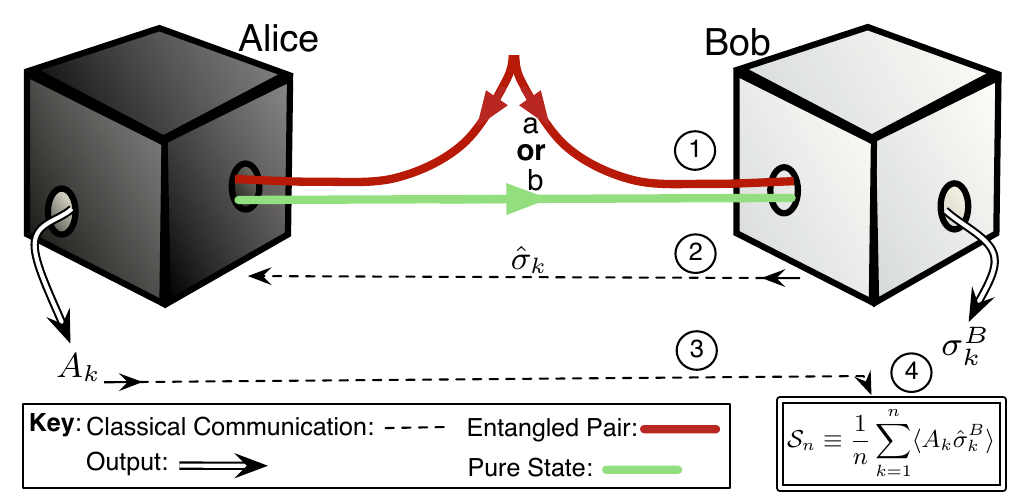}
\end{center}
\vspace{-3ex} \caption{{\bf The steering task.} Bob is skeptical that Alice can remotely affect (\emph{steer}) his state. Bob trusts quantum mechanics (represented by the white box), but makes no assumptions about Alice (represented by the black box). The steps in the task, from top (1) to bottom (4) are as follows. 1. Bob receives his qubit. He is unsure whether he has received half of an  entangled pair (a) or a pure state sent by Alice (b). 2. \emph{After} Bob receives his qubit, he announces to Alice his choice of measurement setting from the set $\{\hat{\sigma}_k^B\}$. 3. Bob records his own measurement results $\sigma_k^B$ and receives the result $A_k$ which Alice declares. 4. Bob combines the results to calculate (over many runs) the steering parameter $\Str_n$. If this is greater than a certain bound, Alice has demonstrated \emph{steering} of Bob's state, and thus Bob can be sure that he received (a) not (b).} \label{fig:conceptual}\end{figure}

EPR-steering is a form of quantum nonlocality strictly weaker \cite{Wiseman2007,JonWisDoh07} than Bell-nonlocality \cite{BelPHY64}, in that one party, Bob, trusts quantum mechanics to describe his own measurements, but makes no assumptions about the distant party, Alice, who has to convince him that she can affect the nature of his quantum state by her choice of measurement setting. So termed in analogy to Bell inequalities, EPR-steering inequalities \cite{CavJonWisRei09} are a superset of the former. Steering inequalities are, in principle, easier to violate experimentally than Bell inequalities because of the asymmetry between the parties; see Fig.~\ref{fig:conceptual}. Thus instead of considering correlation functions for classical variables (measurement outcomes) on the two sides, in EPR-steering one considers correlations between classical variables declared by Alice but quantum expectation values found by Bob. Here we consider {\em linear} EPR-steering inequalities \cite{CavJonWisRei09} involving  statistics collected from an experiment with $n$ measurement settings for each side. For qubits, we can take Bob's $k$th measurement setting to correspond to the Pauli observable $\hat{\sigma}_{k}^B$, along some axis ${\bf u}_{k}$. Denoting Alice's corresponding declared result  (we make no assumption that it is derived from a quantum measurement) by the random variable $A_{k} \in \{-1,1\}$ for all $k$, the EPR-steering inequality is of the form 
\beq
 \Str_n\equiv \frac{1}{n}\sum_{k=1}^n \langle{A_k}\hat{\sigma}_k^B \rangle\leq C_n.\label{GeneralCriterion} 
\eeq

We call the quantity $\Str_n$ the \textit{steering parameter} for $n$ measurement settings. The bound $C_n$ is the maximum value $\Str_n$ can have if Bob has a pre-existing state known to Alice, rather than half of an entangled pair shared with Alice.  It is easy to see that this bound is

\beq
C_n = \max_{\{A_k\}}  \cu{ \lambda_{\rm max} \ro{\frac{1}{n}\sum_{k=1}^n
{A_k}\hat{\sigma}_k^B}}\label{bound}
\eeq
 where $\lambda_{\rm max}(\hat O)$ denotes the largest eigenvalue of $\hat O$.

\begin{figure}
\begin{center}
\includegraphics[width=.94\linewidth]{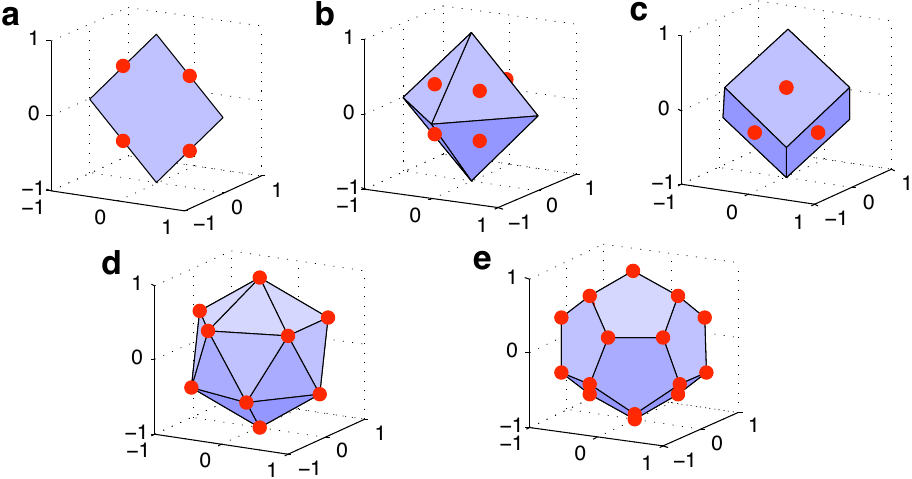}
\end{center}
\vspace{-3ex}
\caption{{\bf Platonic solid measurement schemes.} Measurement axes ${\bf u}_k$ are defined by the Bloch-space directions through antipodal pairs of vertices of regular figures for $n=2,3,4,6$ and $10$ measurements. The figures are: (a) Square, $n=2$; and the four suitable Platonic solids:  (b) Octahedron, $n=3$; (c) Cube, $n=4$; (d) Icosahedron, $n=6$; and (e) Dodecahedron, $n=10$. The {\red$\bullet$} symbols show the orientations of pure states in optimal cheating ensembles for two-qubit Werner states. In (d) and (e) these states align with the measurement axes ({\em vertices}), but in (a), (b) and (c),  they have the {\em dual} arrangement, on the face-centres, similar to the situation in random access codes \cite{spekkensPRL}.}
\label{combinedfigures}\end{figure}

To derive useful inequalities we consider measurement settings based around the four Platonic solids that have vertices that come in antipodal pairs (Fig.~\ref{combinedfigures}). Each pair defines a measurement axis ${\bf u}_k$, giving us 
an arrangement for $n=3$, $4$, $6$, and $10$ settings. For $n=2$ settings, we use a square arrangement. For each measurement scheme (except for $n=10$, which we did not implement experimentally) we do  the following:
\begin{enumerate} 
\item Derive the bound $C_n$ in the inequality (\ref{GeneralCriterion}).\vspace{-1ex}
\item Experimentally demonstrate EPR-steering by violating the inequality using Werner states.\vspace{-1ex}
\item Theoretically show that Alice can saturate the inequality by sending Bob pure states drawn by her from a particular ensemble.   \vspace{-1ex}
\item Experimentally demonstrate 3.~above by nearly saturating the EPR-steering inequality in that way.
\end{enumerate}

\begin{figure}[bp]
\begin{center}
\includegraphics[width=.94\linewidth]{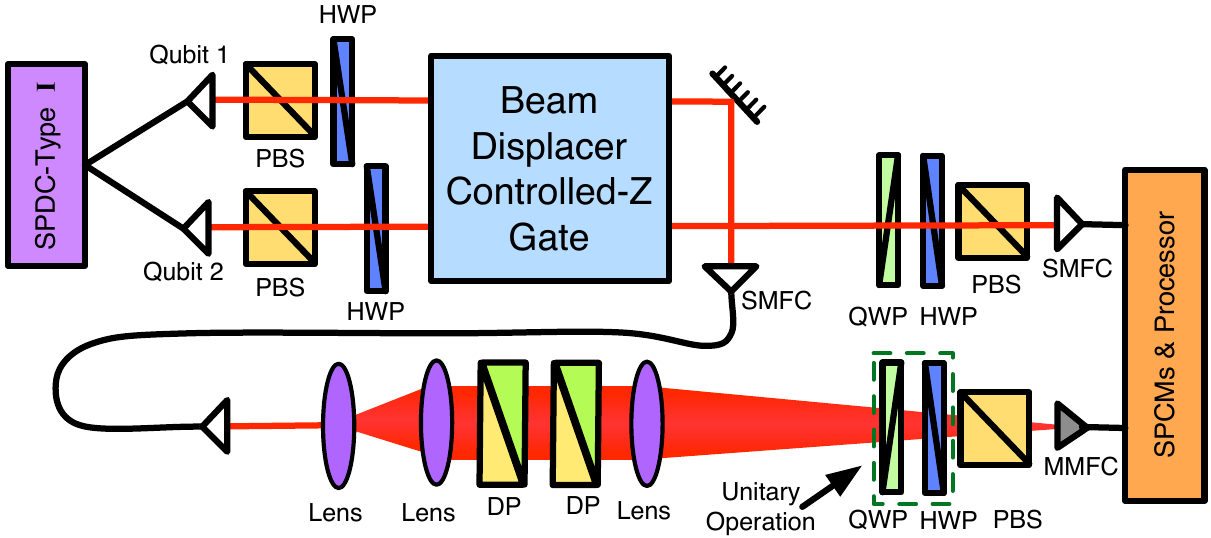}
\end{center}
\vspace{-5ex}
\caption{{\bf Summary of the experimental apparatus.} Pairs of identical photons are produced via type-I SPDC and are confined spatially using single mode fibres (SMFs), which direct them to a linear-optics controlled-Z logic gate. After the gate, qubit 1 is coupled into a SMF passes through a pair of Hanle wedge depolarizers (DPs). By varying the azimuthal angle between the optic axis of the DPs, we control the amount of depolarizing noise, which sets $\mu$.  See Methods for additional details. Abbreviations: PBS: polarizing beam splitter, HWP: half-wave plate, QWP: quarter-wave plate, SMFC: singlemode fibre coupler, MMFC: multimode fibre coupler, SPCM: single photon counting module.}\label{fig:experimental_setup}\end{figure}

\begin{figure*}
\begin{center}
\includegraphics[width=.9\linewidth]{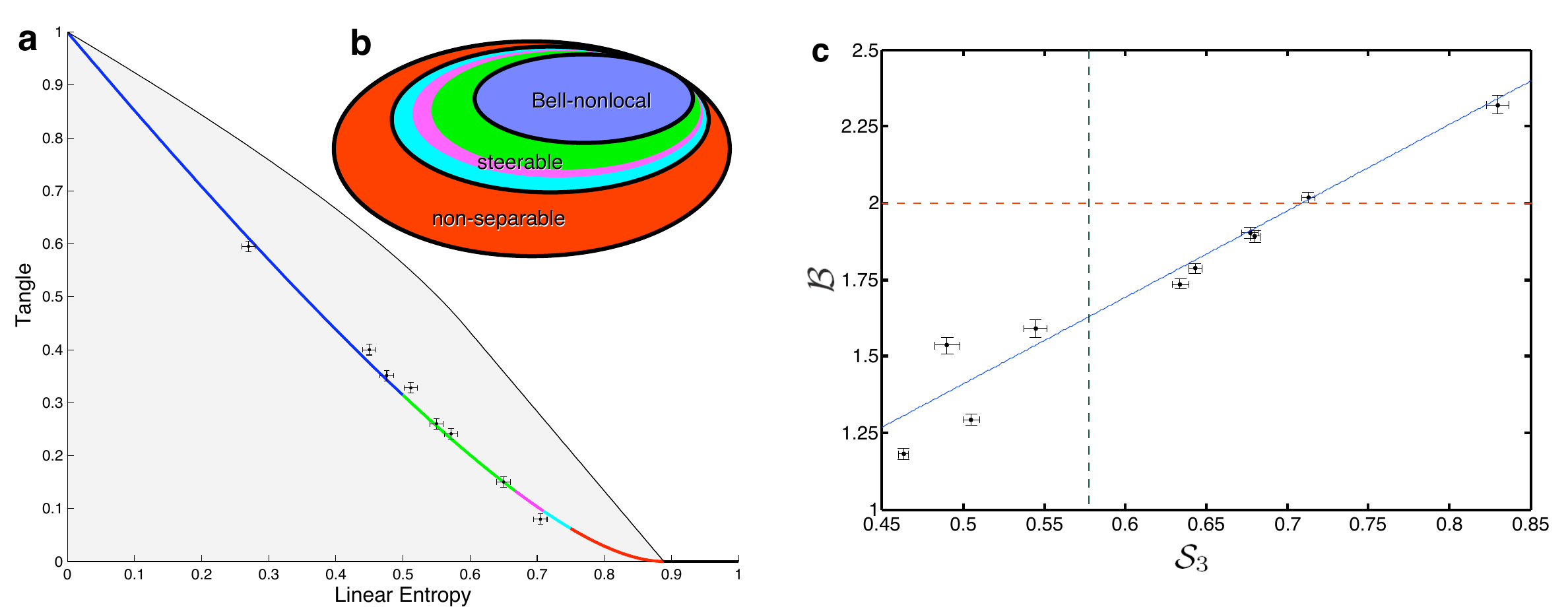}
\end{center}
\vspace{-5ex} \caption{{\bf Experimental demonstration of $\Str_3$-steering of Bell-local states.} \textbf{a.} Tangle-entropy plane \cite{Whiteetal:2007} characterization of the Werner states used for the $\Str_3$ measurements. The thick line represents the theoretical values for the two-qubit Werner state for varying $\mu$. The blue segment represents Werner states able (theoretically) to violate the $\Str_2$ and Bell-CHSH inequalities. The green segment represents Werner states able to violate the $\Str_3$ and $\Str_4$, but not the $\Str_2$, inequalities. The magenta represents those that violate the $\Str_6$ inequality  only, while the cyan represents those steerable with more than 6 settings. Finally, the red region represents Werner states that are entangled but nevertheless not steerable even with infinitely many measurement settings. The black points are experimentally produced states, with tangle and linear entropy calculated from a maximum-likelihood tomographic reconstruction, following ref. \cite{James:2001}.  The error-bars represent one standard deviation calculated from Monte-Carlo generated statistics based on the known Poisson counting distribution. \textbf{b.} Venn diagram conceptually showing the relationship between the different sets of entangled states. Colours are used in the same way as in \textbf{a}, except that the Bell non-local states are those that violate any Bell inequality. \textbf{c.} Experimental demonstration of the hierarchy of Bell-CHSH and $\Str_3$ entanglement regimes. The horizontal red line is the Bell-CHSH inequality bound, and the vertical green line is the $\Str_3$ bound. The diagonal blue curve shows the predicted values for Werner states of varying $\mu$. The data points are experimentally recorded states. The errors bars are generated by propagating the Poissonian noise on the measurement outcomes.}\label{fig:S3_Bell}
\end{figure*}

Werner states \cite{WerPRA89} are the best-known class of mixed entangled states. For qubits, they  can be written as
 \beq 
 W_\mu=\mu\ket{\Psi^-}\bra{\Psi^-}+\left({1-\mu}\right){\mathbf{I}}/{4},\label{Werner2} 
 \eeq 
where $\ket{\Psi^-}$ is the singlet state and $\mathbf{I}$ is the identity, and where $\mu\in[0,1]$. Werner states are entangled iff (if and only if) $\mu> 1/3$ (Ref. \cite{WerPRA89}). They can violate the Clauser, Horne, Shimony and Holt \cite{CHSH:1969} (Bell-CHSH) inequality only if  $\mu > 1/\sqrt{2}$, and cannot violate \textit{any} Bell inequality if $\mu < 0.6595$; Ref. \cite{AciEtalPRA06}. Ref.~ \cite{Wiseman2007} showed that these states are also steerable, with $n\rightarrow\infty$ settings, iff $\mu >  1/2$. With $n=2$ projective measurements they are steerable iff $\mu> 1/\sqrt{2}$, no better than the Bell-CHSH inequality. Deriving analytical expressions for the bounds $C_n$ is a simple exercise in geometry. For the square, octahedron, and cube we find $C_2=1/\sqrt{2}$ and $C_3=C_4= 1/\sqrt{3}\approx 0.5773$. For higher $n$ the exact expressions are lengthy; the approximate numerical values are $C_6 \approx 0.5393$ and $C_{10} \approx 0.5236$. For a Werner state experiment, the expected value of $\Str_n$ is $\mu$ (see below). Thus, using $n\geq 3$ allows us to demonstrate EPR-steering for some Bell-local states, i.e.\ states with $0.6595>\mu>1/2$. Also, with $n$ as small as $6$, $C_n$ is already within $8\%$ of the $n\rightarrow\infty$ limit. 

Consider the EPR-steering experiment, Fig.~\ref{fig:conceptual}, from the point of view of an honest Alice, who does share a suitable entangled state with Bob. She claims to be able to prepare different types of states for Bob by making different remote measurements on her half of the state. If the state is a Werner state, she would claim to be able to prepare mixed states aligned (or anti-aligned) along any Bloch-sphere axis ${\bf u}$. They agree to test this along a specific set of axes $\{ {\bf u}_{k}\}$. To maximize the correlation $\Str_n$  in \erf{GeneralCriterion}, Alice measures $-\s{k}$, and announces her result $A_k$. The value of the correlation $\Str_n$ thus obtained will be $\mu$, independent of $n$, due to the $\hat{U}\otimes\hat{U}\dg$ invariance  of the Werner state. Thus, for a given $n$,  it should be possible to demonstrate EPR-steering if $\mu > C_n$.

We experimentally demonstrate EPR-steering with Werner states in a polarization-encoded two-qubit photonic system. To realize the Werner states, we start by generating identical single photons via type-I spontaneous parametric downconversion (SPDC). These photon pairs are initially unentangled in polarization. We use a non-deterministic controlled-Z (CZ) gate \cite{Obrien:2003,LangfordetalPRL95,KieseletalPRL95,OkamotoetalPRL95} to entangle them in polarization. Ideally, this creates the state $\left(H_1\otimes\mathbf{I}\right)\ket{\Psi^-}$ where $H_1$ is the Hadamard gate acting on qubit 1. Mixture was controllably added, enabling a change of $\mu$, using the depolarizer (DP) method of Puentes \emph{et al} \cite{Puentes:2006}---see Fig.\ref{fig:experimental_setup}. This method produces ``Werner-like'' states---states locally equivalent to the Werner states of \erf{Werner2}.

We implement single-qubit polarization measurements on each qubit using quarter- and half-wave plates (QWPs and HWPs, respectively), a polarizing beam splitter (PBS) and fibre-coupled single photon counting modules (SPCMs), allowing us to measure along arbitrary axes on the Bloch sphere for each qubit. By choosing different combinations of measurement axes, we can perform a variety of measurement tasks: evaluating the Bell-CHSH inequality, evaluating the steering parameter $\Str_n$ for different $n$; or tomographically reconstructing the Werner-like states, $\rho$. As the quantum states $\rho$ are described by $\rho=(\hat{U}\otimes\mathbf{I})W_\mu(\hat{U}\otimes\mathbf{I})^\dagger$, where $\hat{U}$ is a single-qubit unitary operation, we can undo $\hat{U}$ to retrieve a Werner state by incorporating a unitary transformation in the measurement settings of qubit 1.

Figure 4(a) characterizes the Werner states produced in terms of their {\em tangle} and {\em entropy} \cite{Whiteetal:2007}, in relation to the theoretical bounds for nonseparability, steering and Bell-CHSH inequality violation. For each of these states, we measure the Bell-CHSH parameter, $\mathcal{B}$, following Ref. \cite{Kwiat:1999}, and the $\Str_3$ parameter. In Fig.~4(c) we identify several regimes: a region where states violate both a Bell-CHSH inequality and an $\Str_3$ inequality; a region where states violate an $\Str_3$ inequality but \textit{not} a Bell-CHSH inequality; and a region where states violate neither inequality, but are still entangled.

\begin{figure}
\begin{center}
\includegraphics[width=.96\linewidth]{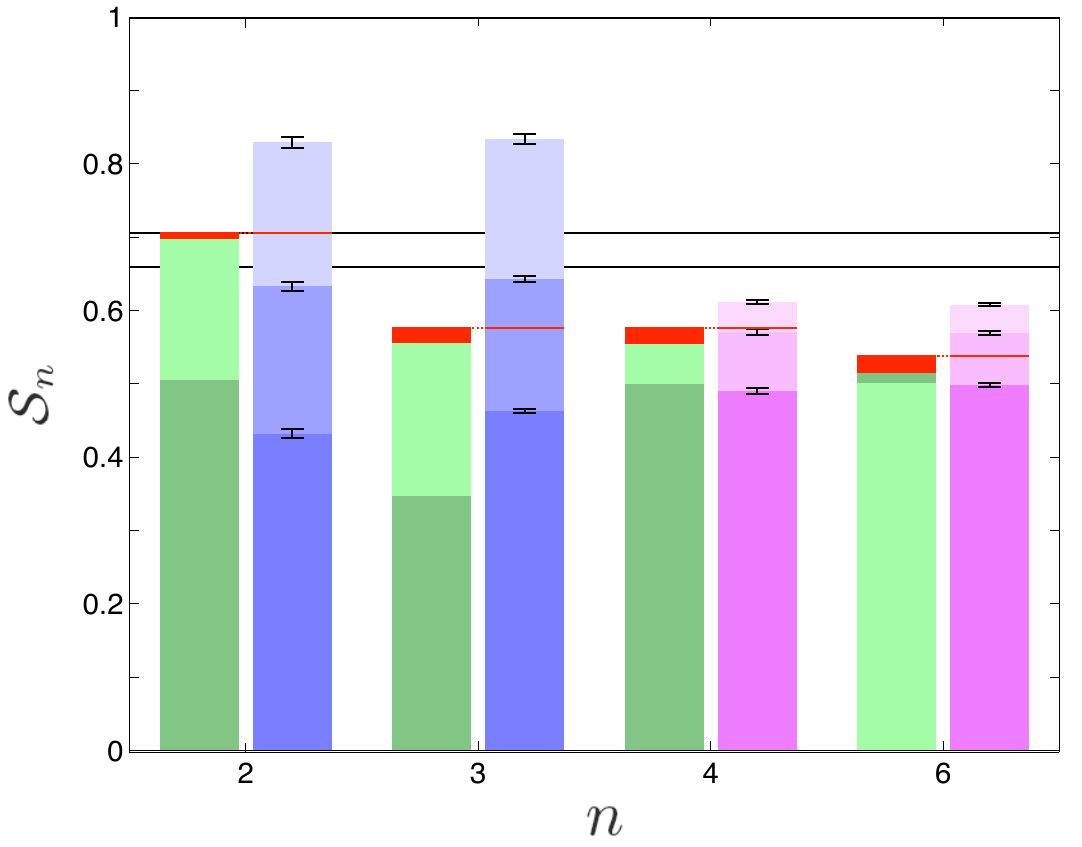}
\end{center}
\vspace{-5ex}
\caption{{\bf Cheating strategies and EPR-steering for increasing $n$.} The top of each bar in the left-hand column, for each $n$, shows the theoretical value of $C_n$ (red), and the experimentally measured values of $\Str_n$ for the vertex- (dark green) and dual- (light green) ensemble cheating strategies. Note that the vertex ensemble is optimal for $n=6$, and that the experimentally measured $\Str_n$ for the optimal cheating strategy nearly saturates the theoretical bound. The tops of the blue and magenta bars represent the measured values of $\Str_n$ for six experimentally-produced Werner states lying near the steering bounds. Bars of the same colour represent the same Werner state measured with different $n$. The mid-blue state demonstrates that there exist Werner states that do not violate a $\Str_2$ inequality but do violate the $n=3$ inequality. Similarly, the mid-magenta state demonstrates the same effect for $n=4$ and $n=6$. Generally good agreement is found between the values of $\Str_n$ measured on the same state, although the state preparation is slightly worse for low values of $\mu$. The error bars are one standard deviation, calculated from Poissonian counting statistics. Error bars not shown are too small to be clearly seen on this scale. The two black lines represent Bell-locality bounds for ideal Werner states with $\Str_n=\mu$: the upper line is the CHSH bound, and the lower line is the value below which states are \textit{unequivocally} Bell-local \cite{AciEtalPRA06}.}
 \label{fig:NvEta}\end{figure}

The amount of entanglement required to demonstrate steering decreases as the number of equally-spaced measurement axes increases, i.e.\ for Platonic solids of increasing order (Fig.~\ref{combinedfigures}). We measure $\Str_n$ for states near the various steering bounds; see Fig.~\ref{fig:NvEta}. We compare the values of $\Str_2$ and $\Str_3$ for each of three particular states ($\mu\approx 0.84$, $\mu\approx 0.67$, $\mu\approx 0.45$) and show that there exist cases where a state violates both the $\Str_2$ and $\Str_3$ inequalities, neither inequality, or---most interestingly---violates the $\Str_3$ but not the $\Str_2$ inequality. Similar behaviour is observed for states ($\mu\approx 0.61$, $\mu\approx 0.57$, $\mu \approx 0.49$) near the $\Str_4$ and $\Str_6$ bounds. The $\Str_3$ and $\Str_4$ comparison is not especially interesting as $C_3=C_4$.

Now consider the the EPR-steering experiment, Fig.~\ref{fig:conceptual}, from the point of view of a {\em dishonest} Alice, who shares no entanglement with Bob. Such an Alice can adopt the following ``cheating'' strategy: Draw a state $\ket{\phi_j} $ from some local-hidden-state (LHS) ensemble $E_n = \cu{\ket{\phi_j}}$ and  send it to Bob. Then, when Bob announces the measurement axis ${\bf u}_k$, announce a result $A_k(j)$ based on this and her knowledge of Bob's state. Although we call this a cheating strategy, Alice cannot actually cheat; the bound $C_n$ in \erf{GeneralCriterion} is defined exactly so that it is saturated by the optimal cheating LHS ensemble. That is, the bounds we have derived are \textit{tight}; a value of  $\Str_n$  greater than $C_n$ is {\em necessary} to demonstrate EPR-steering.

From the symmetry of Bob's measurement scheme, there are two obvious candidate LHS ensembles $E_n$: the vertex-ensemble and the dual-ensemble. In the first case the states $\ket{\phi_j}$ are oriented on the Bloch sphere in the directions of the vertices of the figure defining $\{{\bf u}_k\}$. In the second, they are oriented in the direction of the vertices of the dual figure. Interestingly, both of these  possibilities are optimal, but for different values of $n$; see Fig.~\ref{combinedfigures}. Given an optimal ensemble, Alice's optimal ``cheating'' strategy, having been told Bob's measurement axis ${\bf u}_k$, is to announce as $A$ the {\em more likely outcome} of Bob's measurement on the state $\ket{\phi_j}$ she has sent.

The experimental realization is simple---Alice prepares a single qubit state using a PBS, a HWP and a QWP, and this state is measured by Bob as before. We experimentally demonstrate the near-saturation of the bound $C_n$ using the optimal cheating ensemble for Alice, achieving above $95\%$ saturation for all tested $C_n$, as shown in Fig.~\ref{fig:NvEta}. The small discrepancy from perfect saturation is due to imperfect state preparation and measurement. It is interesting to note that the sensitivity to imperfections increases with $n$, even though the prepared states were tomographically measured to overlap with the ideal states to $> 99\%$ fidelity.

Our demonstration of EPR-steering using states that violate no Bell inequality is possible only because we have broken the conceptual shackles of previous EPR experiments \cite{OuPereira1992,BowenSchnabel2003,HaldSorensen1999,HowellBennink2004}. These followed the approach of Ref.~\cite{ReiPRA89} based on the uncertainty relation for two observables with continuous spectra. We used discrete measurements on entangled qubits, and used up to $n=6$ measurement settings, showing that  that increasing $n$ makes the EPR-steering inequality much {more robust} to noise. Our results open the door to the application of EPR-steering phenomena for nonlocal quantum information processing.

As the degree of correlation required for EPR-Steering is smaller than that for violation of a Bell inequality, it should be correspondingly easier to close the detection loophole and achieve a loophole-free test of steering. This would provide an important and exciting extension of the fundamental principles we have demonstrated.

\begin{acknowledgments}
This work was supported by the Australian Research Council. We thank E. Cavalcanti, A. Fedrizzi, D. Kielpinski, and A.G White for helpful discussions. The authors declare no competing financial interests in this work.
\end{acknowledgments}

\begin{center}{\bf Methods}\end{center}
{\bf Photon source and CZ gate.} Source: A 60mW, linearly polarized, continuous wave 410nm wavelength laser is used to pump a BiBO (bismuth borate) non-linear crystal to produce pairs 820nm single photons via type-I spontaneous parametric downconversion. With a coincidence window of $3ns$, a coincidence rate of approximately 10,000 counts per second is achieved. CZ Gate: The gate is implemented using a passively stable beam displacer configuration, as in Ref. \cite{Obrien:2003}. We are not concerned with the gate's limited success probability in generating entangled states, as we do not aim to close the detection loophole in this experiment.

{\bf Depolarizer Method.} By varying the azimuthal angle between two quartz-glass Hanle depolarizers \cite{Puentes:2006}, we creates a tunable, variable depolarizing device. It couples the polarization degree of freedom to the spatial degree of freedom---tracing over spatial information induces mixture. By optimizing these procedures, high-quality Werner states (fidelity $\geqslant 94\%$ in each case) can be produced for a wide range of $\mu$. 

{\bf From ``Werner-like'' to Werner.} The Werner-like states $\rho$ produced by the depolarizer method can be rotated to Werner states (Eq.~\ref{Werner2}) with a single-qubit unitary operation $\hat U$.  To calculate the optimal unitary operation, we first tomographically reconstruct $\rho$ following Ref. \cite{James:2001}. We numerically search for $\hat{U}$ by minimizing: $\text{Cost}=1-\text{fidelity}\left( (\hat{U}\otimes\mathbf{I})\rho(\hat{U}^\dagger\otimes\mathbf{I}),W_\mu\right).$ The $\hat{U}$ so determined is then used to rotate the measurement settings for qubit 1.

{\bf Calculating $C_n$.} In each case we search over the possible sets $\{A_k\}$,
numerically evaluating the maximand in \erf{bound}.   Then, choosing one the sets
that attains the maximum, we use the geometry of the relevant Platonic solid to
evaluate it analytically. The same analytical expressions are found from the optimal LHS
ensembles $E_n$ of Fig.~\ref{combinedfigures}. Those not given in the main text are:
\bqa
C_6&=&  1
-\rfrac{5L_6}{12}\sqrt{4-{\rm sec}^2 \rfrac{3\theta}{2}} \\
C_{10}&=& 1 -\frac{1}{10}\left( 1 +
 \frac{\tan 2\theta} {\sin\theta} \right)\sqrt{9L_{10}^2-4} .
\eqa
Here $L_6=4/(\sqrt{10+2\sqrt{5}})$ and $L_{10}=4/(\sqrt{15}+\sqrt{3})$ are the side lengths of an
icosahedron and dodecahedron respectively, circumscribed by the
Bloch sphere, and $\theta = \pi/5$.

\end{document}